\documentclass[prb,twocolumn,showpacs,preprintnumbers,amsmath,amssymb,superscriptaddress, bibnotes]{revtex4}

\usepackage{color}
\usepackage{graphicx}
\usepackage{dcolumn}
\usepackage{bm}

\begin{document}

\title{Normal-state properties of the antiperovskite oxide Sr$_{3-x}$SnO revealed by $^{119}$Sn-NMR}

\author{Shunsaku~Kitagawa}
\email{kitagawa.shunsaku.8u@kyoto-u.ac.jp}
\author{Kenji~Ishida}
\author{Mohamed~Oudah}
\affiliation{Department of Physics, Kyoto University, Kyoto 606-8502, Japan}
\author{Jan~Niklas~Hausmann}
\affiliation{Department of Physics, Kyoto University, Kyoto 606-8502, Japan}
\affiliation{Department of Chemistry, Humboldt-Universit\"{a}t zu Berlin, Berlin 12489, Germany}
\author{Atsutoshi~Ikeda}
\author{Shingo~Yonezawa}
\author{Yoshiteru~Maeno}
\affiliation{Department of Physics, Kyoto University, Kyoto 606-8502, Japan}

\date{\today}

\newcommand{\red}[1]{\textcolor{red}{#1}}

\begin{abstract}
We have performed $^{119}$Sn-NMR measurements on the antiperovskite oxide superconductor Sr$_{3-x}$SnO to investigate how its normal state changes with the Sr deficiency. 
A two-peak structure was observed in the NMR spectra of all the measured samples. 
This suggests that the phase separation tends to occur between the nearly stoichiometric and heavily Sr-deficient Sr$_{3-x}$SnO phases.
The measurement of the nuclear spin-lattice relaxation rate $1/T_1$ indicates that the Sr-deficient phase shows a conventional metallic behavior due to the heavy hole doping.
In contrast, the nearly stoichiometric phase exhibits unusual temperature dependence of $1/T_1$, attributable to the presence of a Dirac-electron band.
\end{abstract}

\abovecaptionskip=-5pt
\belowcaptionskip=-10pt

\maketitle

Recently, antiperovskite (inverse-perovskite) oxides have attracted much attention because of their nontrivial electronic structures\cite{T.Kariyado_JPSJ_2011,T.H.Hsieh_PRB_2014}. 
It was shown from the investigation with the data-mining algorithm that a number of compounds including antiperovskite oxides possess the inversions of the valence and conduction bands\cite{M.Klintenberg_APR_2014}.
Actually, the band calculations of Ca$_3$PbO and its family indicate that a Dirac cone due to the $p$~-~$d$ band inversion and protected by crystalline symmetry exists around the Fermi energy $E_{\rm F}$ and that the Dirac cone has a slight but nonzero band gap because of the higher-order effect of the spin-orbit coupling\cite{T.Kariyado_JPSJ_2011,T.Kariyado_JPSJ_2012}.
In addition, Hsieh {\it et al.} theoretically revealed that the antiperovskite oxides with the band inversion and the gap opening are topological crystalline insulators\cite{T.H.Hsieh_PRB_2014}. 

\begin{figure}[!t]
\vspace*{10pt}
\begin{center}
\includegraphics[width=8.5cm,clip]{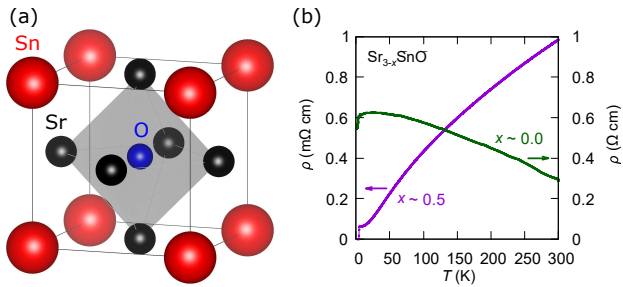}
\end{center}
\caption{(Color online) (a)Crystal structure of the antiperovskite Sr$_{3}$SnO.
The black, red, and blue spheres represent the strontium, tin, and oxygen atoms, respectively.
This figure is prepared with the program VESTA\cite{K.Momma_JAC_2008}.
(b) Temperature dependence of the resistivity at $x \sim 0$ and $x \sim 0.5$.
}
\label{Fig.1}
\end{figure}

From a crystallographic point of view, the antiperovskite oxide $A_3B$O is the charge-inverted counterpart of the ordinary perovskite oxide in which the position of metal and oxygen ions are reversed as shown in Fig.~\ref{Fig.1}(a). 
While the metal element at the unit-cell center is octahedrally surrounded by oxygens in an ordinary perovskite oxide, the oxygen atom is octahedrally coordinated by metal elements in an antiperovskite oxide.
The space group of antiperovskite oxides is basically the same as that of the ordinary cubic perovskite oxides ($Pm$\={3}$m$, No. 221, O$_h^1$)\cite{A.Widera_MRB_1980}. 
When $A$ is a group-2 element and $B$ is tin or lead, the valence of the $B$ ion becomes 4$-$ to satisfy the charge neutrality condition\cite{J.Nuss_ACB_2015}. 
Indeed, Sn$^{4-}$ state was experimentally confirmed in Sr$_{3-x}$SnO by M\"{o}ssbauer measurements\cite{M.Oudah_submit_2018}; such a negative-ionic state of a metallic element is quite unusual.

The predicted Dirac dispersion has been experimentally observed in single-crystalline Ca$_{3}$PbO through soft x-ray angle-resolved photoemission spectroscopy\cite{Y.Obata_PRB_2017}.
Moreover, semiconducting behavior and room-temperature ferromagnetism were observed in thin films of nearly stoichiometric Sr$_{3-x}$SnO ($x \sim$ 0)\cite{Y.F.Lee_APL_2013,Y.F.Lee_MRS_2014,Y.F.Lee_JAP_2014}.
Some of the present authors reported the discovery of the superconducting (SC) transition occurring at $T_{\rm c} = 5$~K on the Sr-deficient antiperovskite oxide Sr$_{3-x}$SnO ($x \sim$ 0.5), which marks the first superconductor among the antiperovskite oxides\cite{M.Oudah_NatCommun_2016}.
Since the possibility of the topological crystalline superconductivity has been theoretically proposed due to the strong $p$~-~$d$ mixing of the orbitals in a moderately hole-doped Sr$_{3-x}$SnO with a rigid-band shift, SC properties on Sr$_{3-x}$SnO have attracted much attention.
Quite recently, some of the present authors systematically studied the Sr deficiency $x$ dependence of the SC state and found that superconductivity with a highest volume fraction of as much as 70\% appears in the vicinity of $x = 0.5$ (sample weight is $\sim$ 30~mg)\cite{M.Oudah_submit_2018}.
In addition, a second superconducting transition was observed at $\sim$ 1~K for all SC samples.
However, microscopic investigations of physical properties of Sr$_{3-x}$SnO have not been reported so far, and this makes the understanding of the normal and SC states difficult.

In this Rapid Communication, we report results of $^{119}$Sn-NMR measurements, which have been performed to investigate the sample dependence of the electronic states of Sr$_{3-x}$SnO with varying $x$ contents.
We measured four samples, and found a two-peak structure in NMR spectrum arising from the main phases, along with a tiny peak from an impurity phase.  
The two-peak structure indicates that the phase separation tends to occur between the nearly stoichiometric and Sr-deficient Sr$_{3-x}$SnO phases.
We also measured the nuclear spin-lattice relaxation rate divided by temperature $1/T_1T$ for the two main phases, and found the electronic states are quite different between the two phases: The nearly stoichiometric phase shows a characteristic behavior of a Dirac-electron system, but the Sr-deficient phase shows a conventional-metal behavior with the $1/T_1T$ value smaller than that of the Sn-metal by a factor of about 4.
These $1/T_1$ results suggest that both 5~K and 1~K superconductivities arise from the Sr-deficient Sr$_{3-x}$SnO with a certain $x$ causing a heavy hole-carrier doping. 

The measured four samples are listed in Table~\ref{Tab.1}.
Polycrystalline samples of Sr$_{3-x}$SnO were synthesized by the reaction of Sr chunk (Furuuch Metal Co., 99.9\%) and SnO powder (Furuuchi Metal Co., 99.9\%) in an alumina crucible sealed inside a quartz tube under vacuum 
or 0.3 bar of argon at room temperature\cite{M.Oudah_NatCommun_2016,J.N.Hausmann_SST_2018,A.Ikeda_PhysicaB_2018,M.Oudah_submit_2018}.
The molar ratios of the starting materials are shown in Table~\ref{Tab.1}.
The sealed quartz tubes were heated to 825$^{\rm o}$C over 3~h and kept at 825$^{\rm o}$C for 3~h. 
Then, the tubes were immediately quenched in water.
The actual values of $x$ in the samples A and C were changed from the starting values due to the evaporation of strontium, while the evaporation in other samples was negligible due to the effect of argon gas.
The actual values of $x$ in the samples A and C were evaluated as $\sim$ 0.5 and $\sim$ 0 from the estimation of the amount of evaporated strontium with the weight change.
Powder x-ray diffraction measurements revealed cubic Sr$_{3-x}$SnO phase to be dominant in all samples.
To prevent sample degradation by oxidization and/or moisture, all polycrystalline samples were mixed with epoxy (Emerson \& Cuming, Stycast 1266), and the mixture was solidified with random crystal orientation. 
All procedures were done in a glove box filled with argon.

Zero-resistivity with the metallic resistive behavior was observed in a similar quality sample as the samples A and B, but not in a similar quality sample as the samples C and D, although the precursor of superconductivity was observed accompanied with semiconducting behavior [Fig.~\ref{Fig.1}~(b)]. 
Note that the samples are not stable enough in air, so that the samples used for the resistive measurements cannot be used for other measurements.
Superconductivity was also confirmed by magnetization measurements with a commercial SQUID magnetometer (Quantum Design, MPMS) after mixing with Stycast.
The SC volume fractions in the samples A and B were estimated as $6 - 8$\%.
A large sample volume to obtain enough NMR signal ($\sim$ 300~mg) prevents high SC volume fraction.
A conventional spin-echo technique was used for the NMR measurements.
The $^{119}$Sn-NMR spectra (nuclear spin $I$ = 1/2, nuclear gyromagnetic ratio $^{119}\gamma/2\pi = 15.8775$~MHz/T, and natural abundance 8.6\%) were obtained as a function of magnetic field in a fixed frequency $f$ = 99.1~MHz.
Since the applied magnetic field is $\sim$ 6.2~T, which is much higher than the upper critical field $\mu_0 H_{\rm c2} \sim 0.5$~T, we could not investigate SC properties of Sr$_{3-x}$SnO in the present study.
The $^{119}$Sn Knight shifts of the samples were calibrated using a Sn metal, and $1/T_1$ of $^{119}$Sn was determined by fitting the time variation of the spin-echo intensity after the saturation of the nuclear magnetization to a single exponential function.
All recovery curves can be fit by a unique $T_1$ value as shown in the Supplemental Material.

\begin{table}[tp]
\caption[]{Synthesis conditions and basic properties of the Sr$_{3-x}$SnO samples.} 
\vspace{0.5cm}
\begin{tabular}{ccccc}
\hline
sample & \shortstack{starting \\ molar ratio \\Sr : SnO} & \shortstack{synthesis \\ pressure} &  $x$ & \shortstack{super- \\ conductivity}  \\
\hline
A & 3.0 : 1 & vacuum & $\sim$ 0.5 & Yes \\
B & 2.5 : 1 & 0.3 bar (Ar) & $\sim$ 0.5 & Yes \\
C & 3.5 : 1 & vacuum & $\sim$ 0 & No \\
D & 3.0 : 1 & 0.3 bar (Ar) & $\sim$ 0 & No \\
\hline
\end{tabular}
\label{Tab.1}
\end{table}

\begin{figure}[!tb]
\vspace*{10pt}
\begin{center}
\includegraphics[width=8.5cm,clip]{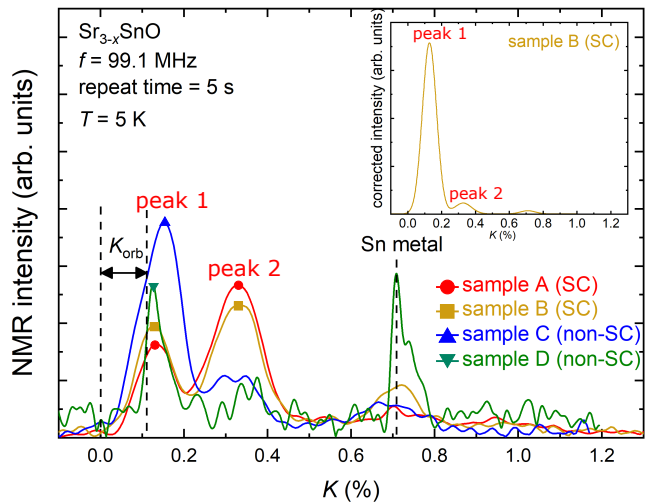}
\end{center}
\caption{(Color online) $^{119}$Sn Knight-shift spectra at $T = 5$~K and $f = 99.1$~MHz of all measured samples in their normal state converted from field-swept NMR spectra.
The repetition time of the RF pulse fields is 5 seconds, which is shorter than $T_1$ at the peak 1.
Thus, the intensity of the peak 1 underestimates the corresponding volume fraction.
The intensity of each spectrum is normalized by the total area.
The positions of $K$ = 0\%, $K_{\rm orb}$ = 0.11\%,  and $K$ for the Sn metal ($K$ = 0.717\%) are indicated by broken lines.
The labels SC and non-SC indicate whether the sample shows superconductivity in SQUID magnetometor.
(Inset) Estimated NMR spectrum of sample B whose intensity of the peak 1 is corrected based on the value of $T_1$.
}
\label{Fig.2}
\end{figure}

Figure~\ref{Fig.2} shows the Knight-shift spectra at $T = 5$~K of all measured samples, which were converted from field-swept NMR spectra.
In the measurements, the repetition time of the RF pulse fields was 5 s, which is much shorter than the $T_1$ of the stoichiometric phase but substantially longer than that of the Sr-deficient phase as discussed later.
Thus, it should be noted that the intensity of these spectra does not reflect the volume fraction of the samples. 
The intensity of each spectrum obtained in the measurement was normalized by the total area.
In the field-swept NMR spectrum, a resonance magnetic field $\mu_0 H_{\rm res}$ is shifted as a function of the Knight shift $K$, and $K$ is linearly proportional to the spin susceptibility $\chi$ at the nuclear site as
\begin{align}
K &= \frac{H_0-H_{\rm res}}{H_{\rm res}}\\
&= K_{\rm spin} + K_{\rm orb} = A_{\rm hf}\chi + K_{\rm orb},
\end{align}
where $H_0$ is a magnetic field at $^{119}K = 0$, $A_{\rm hf}$ is the hyperfine coupling constant, $K_{\rm spin} = A_{\rm hf}\chi$ is the spin part of the Knight shift and $K_{\rm orb}$ is the temperature independent orbital part of the Knight shift.
$K_{\rm orb}$ is estimated to be 0.11\% from the relationship between the Knight shift and $1/T_1T$; $1/T_1T = K_{\rm spin}^2$.
It was shown that all spectra contains some NMR signals around the $K$ position for the Sn-metal ($K = 0.717$\%)\cite{Metallicshifts_1977}, which originates from an impurity phase and/or the decomposition of the samples.
It should be noted that this peak is strong only in the sample D, which was measured six months after the mixing with Stycast while other samples were measured just after the mixing.
This indicates that the decomposition occurs for a longer term even in the powder sample embedded in Stycast.
Except for the signals due to the Sn metal, a two-peak  structure was observed in all samples, indicative of the phase separation occurring in all four samples.
Although the precise discussion about the fraction of each phase cannot be made from Fig.~\ref{Fig.2}, the NMR intensity ratio of the two peaks was similar between the samples A and B or between the samples C and D with similar $x$ contents; the ratio was different between the samples with different Sr deficiencies.
Since the intensity of the peak 1 is larger in the nearly stoichiometric samples C and D, we identify the peaks 1 and 2 as arisng from the nearly stoichiometric and Sr-deficient Sr$_{3-x}$SnO phases, respectively.

\begin{figure}[!tb]
\vspace*{10pt}
\begin{center}
\includegraphics[width=8.5cm,clip]{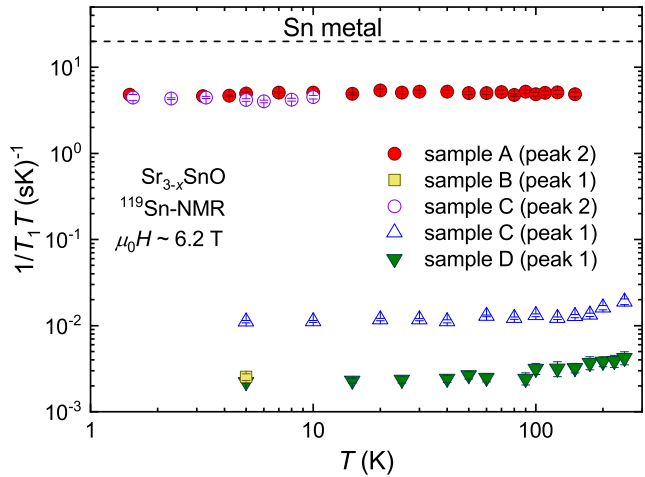}
\end{center}
\caption{(Color online) Temperature dependence of $1/T_1T$ at different peak positions and samples on a logarithmic scale.
$1/T_1T$ is measured around 6.2~T.
The broken line represents the value for Sn metal.
Since $1/T_1T$ is proportional to the square of the electronic density of states (DOS), peak 2 corresponds to a conventional metallic phase and peak 1 corresponds to a less metallic phase with a quite low DOS.
}
\label{Fig.3}
\end{figure}

To investigate electronic properties of each phase, we measured $1/T_1$ at the two peaks.
The temperature dependencies of $1/T_1T$ at the two peaks are shown in Fig~\ref{Fig.3}.
$1/T_1T$ at the peak 2 is independent of temperature between 1.5~K and 150~K, which is a typical behavior of a conventional metallic system, and the value of $1/T_1T$, proportional to the square of the electronic density of states (DOS), was about 1/4 of that of the Sn metal.
Thus, we conclude that the peak 2 originates from a conventional metallic phase.
The value of $1/T_1T$ at the peak 2 in the sample C was identical to that in the sample A, indicating that the microscopic $x$ content of the peak 2 is the same in both samples.
$1/T_1T$ at the peak 1 arising from the nearly stoichiometric phase in the samples C and D shows a temperature independent behavior with a very small value of $1/T_1T$ up to 150 K and a superlinear behavior above 150 K.
Similar behavior was observed in three-dimensional Dirac and Weyl semimetal systems such as Sr$_3$PbO\cite{K.Kitagawa_JPS_2017}, PtSn$_4$\cite{Y.Nakai_JPS_2017}, and TaP\cite{H.Yasuoka_PRL_2017}.
The robust temperature independent $1/T_1T$ at low temperatures indicates that the nearly stoichiometric phase is not in the semiconductor state but possesses a finite DOS at $E_{\rm F}$, although resistivity shows weakly semiconducting behavior as shown in Fig.~\ref{Fig.1} (b).
The value of $1/T_1T$ at the peak 1 was 400 times smaller than that at the peak 2.
The small value of $1/T_1T$ indicates a quite low DOS around $E_{\rm F}$ in the nearly stoichiometric phase.

On warming, $1/T_1T$ increases above 150~K, which is consistent with a Dirac-electron property.
In the case of a pure Dirac point exactly located at $E_{\rm F}$ with linear $k$ dispersion, $1/T_1T \propto T^4$ ($T^2$) in 3D (2D) systems.
On the other hand, a Dirac point is located slightly above/below $E_{\rm F}$ in most actual compounds.
Therefore, the value of $1/T_1T$ is affected when the temperature is higher than the energy difference between $E_{\rm F}$ and the Dirac point.
In fact, such a temperature dependence was observed in the Weyl-semimetal TaP: $1/T_1T$ is nearly constant below 10 K and shows $T^{2}$ dependence above 30~K\cite{H.Yasuoka_PRL_2017}. 
According to band calculations, it is expected that $E_{\rm F}$ in stoichiometric Sr$_{3}$SnO is located slightly below the gapped Dirac point, which would affect the value of $1/T_1T$ at high temperatures\cite{A.Ikeda_PhysicaB_2018}.

\begin{figure}[!tb]
\vspace*{10pt}
\begin{center}
\includegraphics[width=8.5cm,clip]{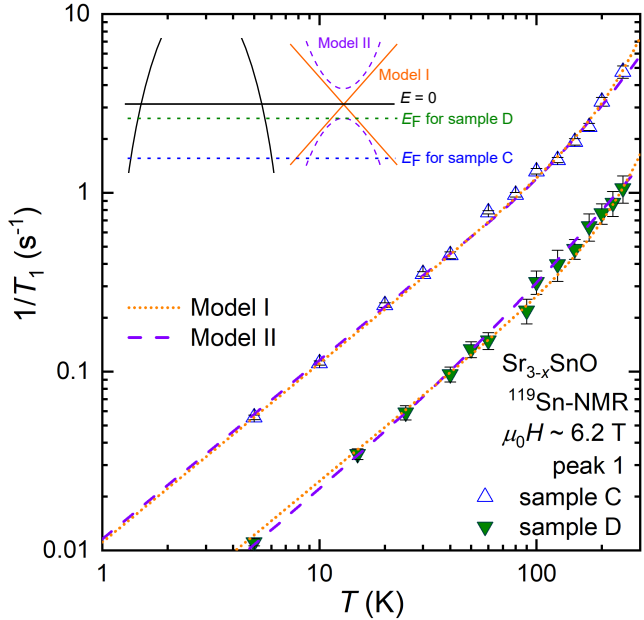}
\end{center}
\caption{(Color online) Temperature dependence of $1/T_1$ calculated by two models. 
Triangles represent the observed values of $1/T_1$ measured at the peak 1 in the samples C and D.
(Inset) Schematic band dispersions in the two models.
}
\label{Fig.4}
\end{figure}

To explain the temperature dependence of $1/T_1$ at the peak 1, we adopt two-channel relaxation models with different energy dependencies of the DOS as follows,
\begin{align*}
\frac{N(E)}{N(0)} = \begin{cases}
1 + a E^2 \hspace{45pt} (\mathrm{model~I}),\\
1 + b \sqrt{|E| - \frac{\Delta}{2}} \hspace{14pt} (\mathrm{model~II} ,|E| > \Delta/2),\\
1 \hspace{75pt} (\mathrm{model~II}, |E| \leq \Delta/2),
\end{cases}
\end{align*}
where the constant component arises from a finite DOS of a conventional hole band, $\Delta$ is the energy gap of split bands in the model II.
We set the zero energy to the Dirac point (model I) or the center of the band gap (model II).
The linear ($E \propto k$) and parabolic ($E \propto k^2$) dispersions of the bands are assumed in model I and II, respectively.
Schematic images of two models are illustrated in the inset of Fig.~\ref{Fig.4}.
The relaxation rate $1/T_1$ in the metallic state can be calculated as,
\begin{align*}
1/T_1 \propto \int^{\infty}_{-\infty}N(E)^2 f(E)[1-f(E)] dE,
\end{align*}
where $f(E) = 1/\{\exp[(E-E_{\rm F})/k_{\rm B}T]+1\}$ is the Fermi distribution function.
As shown in Fig.~\ref{Fig.4}, the temperature dependences of $1/T_1$ are consistently reproduced by the two models.
The parameters used in the calculations are listed in Table~\ref{Tab.2}.
It is noted that three times difference of the $b$ coefficient is not significant since the temperature dependence of $1/T_1$ is not so sensitive against the $b$ coefficient.
Actually, $1/T_1$ can also be reproduced by a gapped Dirac model. 
We cannot distinguish which model is more appropriate for the temperature dependence of $1/T_1$ of the nearly stoichiometric Sr$_{3-x}$SnO below room temperature.
To reveal topological properties of nearly stoichiometric Sr$_{3-x}$SnO, further experiments with samples free from the Sr deficiency are needed.

Next, we discuss how the ground state changes from the non-SC to SC samples with the amount of the Sr deficiency.
The present NMR results indicate that the amount of the Sr deficiency does not change continuously but is stabilized at certain values discontinuously.
It seems that it is difficult to synthesize a sample with high SC volume fraction since there is a miscibility gap in the the nearly stoichiometric and Sr-deficient Sr$_{3-x}$SnO phases.
The careful investigation of sample properties and the $x$ dependence will be reported by some of the authors\cite{M.Oudah_submit_2018}.
Based on our NMR measurements, the peak 2 originates from a conventional metallic Sr$_{3-x}$SnO phase with the heavy hole-doping due to the Sr deficiency, and this phase would show superconductivity.
Note that two SC transitions at 5~K and 1~K are reported in Sr$_{3-x}$SnO ($x \sim 0.5$).
Since the two SC transitions are always observed together in one sample, both seem to originate from Sr-deficient Sr$_{3-x}$SnO.
According to the band calculations, $E_{\rm F}$ of the heavily hole-doped phase is far from the Dirac point\cite{A.Ikeda_PhysicaB_2018}.
On the other hand, the peak 1 originates from the less metallic Sr$_{3-x}$SnO phase with a quite low DOS.
Interestingly, the value of $1/T_1T$ at the peak 1 in the sample C was slightly larger than that in the sample D, indicating that the doping level is different between these samples.
In fact, Knight shift at the peak 1 in the sample C is slightly larger than that in the sample D, which is consistent with the results of $1/T_1T$ measurements.

\begin{table}[!tb]
\vspace*{1cm}
\caption[]{Parameters used in the calculation of $1/T_{1}$}
\vspace*{0.5cm}
\begin{tabular}{cccccc}
\hline
sample & model & $a$(J$^{-2}$) & $b$(J$^{-0.5}$) & $E_{\rm F}/k_{\rm B}$ (K) & $\Delta/k_{\rm B}$ (K)  \\
\hline
C & I & 1.2$\times10^{-6}$ & - & -200 & - \\
  & II& - & 0.09 & -200 & 100 \\
D & I & 1.2$\times10^{-6}$ & - & -50 & -\\
  & II& - & 0.03 & -50 & 100 \\
\hline
\end{tabular}
\label{Tab.2}
\end{table}

We tentatively estimated the volume fraction of each phase from the NMR spectra by taking into account longer $T_1$ of the nearly stoichiometric phase.
After a suitable correction, the ratio of the NMR peak intensities becomes, roughly speaking, proportional to that of each phase in the sample.
By using the NMR intensity corrected with the $T_1$ values, the volume ratios of the nearly stoichiometric and Sr-deficient Sr$_{3-x}$SnO phases in the samples A, B, C, and D were estimated to be 10:1, 11:1, 65:1, and 50:1, respectively.
The NMR spectrum of the sample B, whose intensity of the peak 1 is corrected by taking into account the value of $T_{1}$, is shown in the inset of Fig.~\ref{Fig.2}.
The volume fractions of the SC phase in the samples A and B seem to be consistent with the SC volume fraction estimated from dc magnetization measurements.

In conclusion, we have performed $^{119}$Sn-NMR measurements to investigate how the ground state of Sr$_{3-x}$SnO changes from the non-SC to SC phase against the Sr deficiency.
In the four samples we measured, a two-peak structure of the NMR spectra was observed, indicating that the amount of the Sr deficiency does not change continuously but is stabilized at certain values discontinuously.
The two-peak structure means that the phase separation tends to occur between the nearly stoichiometric and heavily Sr-deficient Sr$_{3-x}$SnO phases.
$1/T_1$ measurements indicate that the nearly stoichiometric phase shows a characteristic behavior of a Dirac-electron system, but the Sr-deficient phase shows a conventional-metal behavior with the $1/T_1T$ value smaller than that of the Sn-metal by a factor of about 4.
The $1/T_1$ result suggests that both 5~K and 1~K superconductivities arise from the Sr-deficient Sr$_{3-x}$SnO with a certain $x$ with heavy hole-carrier doping. 

\section*{Acknowledgments}
We would like to acknowledge M. Maesato for technical support. 
This work was partially supported by the Kyoto Univ. LTM Center, Grant-in-Aids for Scientific Research (KAKENHI) (Grant Nos JP15H05882, JP15H05884, JP15K21732, JP25220710, JP15H05745, JP17K14339, JP15H05851, JP15K21717, and JP17J07577) and by the JSPS Core-to-Core Program (A. Advanced Research Network), as well as by Izumi Science and Technology Foundation (Grant No. H28-J-146). 
A. I. is supported by Japan Society for the Promotion of Science as JSPS Research Fellow.


%

\end{document}